\documentclass[conference]{IEEEtran}
 
\usepackage{optidef}

\ifCLASSINFOpdf
   \usepackage[pdftex]{graphicx}
\else
\fi
\usepackage{amsmath}
\usepackage{amssymb}
\usepackage{algorithmic}

\usepackage{url}

\usepackage[]{fancyhdr} %
\newcommand{\changefont}{\fontsize{9}{9}\selectfont}
\ifCLASSOPTIONcompsoc
 \usepackage[caption=false,font=normalsize,labelfont=sf,textfont=sf]{subfig}
\else
 \usepackage[caption=false,font=footnotesize]{subfig}
\fi

\newcommand{\Pro}{\mathbb{P}}

\IEEEoverridecommandlockouts
\begin{document}
\bstctlcite{BanffConf:BSTcontrol}

%
\title{Sizing Co-located Storage for Uncertain Renewable Energy Sold Through Forward Contracts}

\author{\IEEEauthorblockN{Tomas Valencia Zuluaga and Shmuel S. Oren}
\IEEEauthorblockA{Industrial Engineering \& Operations Research\\University of California at Berkeley\\
tvalenciaz@berkeley.edu, shmuel@berkeley.edu}
}



%





\maketitle
\thispagestyle{fancy}
\pagestyle{fancy}


\begin{abstract}
In this paper, we propose a high-level Stochastic steady-state model to analyze the value of co-located energy storage systems for wind power producers that participate in an electricity market through Forward or Day Ahead contracts.
In particular, we try to find optimal sizing and contracting and stationary operating policies for profit maximization in the long-run.
We obtain a stylized model calibrated to actual wind power production and electricity wholesale price data that allows us to asses the value  of storage size and perform sensitivity analysis on key parameters such as contract prices, storage cost and storage efficiency.     

\end{abstract}



%
\IEEEpeerreviewmaketitle

\section{Introduction}

The impending consequences of climate change has driven massive proliferation of renewable energy  resources (RES) around the world. However, a key obstacle to large-scale integration of RES in power systems is the short-term uncertainty and variability of their output. This poses both technical challenges for the reliable operation of the power system and financial challenges for investors in RES, since it is hard to guarantee a reliable income flow without RES implicit subsidies  through policies such as feed-in tariffs and mandatory contracting for retailers. Such policies however are not sustainable and are being challenged due to their cost to consumers and many systems are moving toward imposing  scheduling requirements and forward commitments on RES forcing them to compete on a level playing field with other resources.  
The addition of Battery Energy Storage Systems (BESS) to renewable power plants can help mitigate their uncertainty and thus can help towards solving both the technical and financial issues.
While it is intuitive to understand how storage can mitigate  uncertainty, sizing and managing BESS is not obvious. 

Optimizing the sizing and operation of energy storage for intermittent power plants has been a very active research topic for several years, and there are numerous references of relevant related work, both theoretical as well as in  applied cases. 
In \cite{kim_optimal_2011}, a theoretical analysis is performed for the case of a wind farm that uses storage to optimize its bidding strategy on the day-ahead market in order to minimize imbalance penalties. In this, and other works that study the  behavior of wind power producers (WPP) in day-ahead markets \cite{bitar_bringing_2012,dent_opportunity_2011}, it is observed that the optimal bidding strategy for WPP takes the form of an optimal fractile, in the spirit of a Newsvendor problem. In fact, the bidding problem of a WPP is a reverse newsvendor problem, where the uncertainty lies not on the  supply side.

There are also numerous references that address the issue of optimal sizing and optimal management of co-located energy storage.
In \cite{harsha_optimal_2015}, the infinite-horizon average cost of electricity purchases is minimized by finding an optimal storage management policy and optimal storage size  for a  power plant that serves a local demand and purchases any shortfall from the grid in presence of dynamic pricing.
The optimality of the balancing storage management policy and the authors prove optimality of a dual threshold policy, in the spirit of optimal $(s,S)$ policies for inventory. 
In \cite{ru_storage_2013}, the optimal size of a BESS is found for a grid-connected PV system that can purchase and sell energy from the grid under time-of-use pricing, and the convexity of profit in storage size is shown.
In \cite{loukatou_optimal_2021}, the value of co-located storage is analyzed for the case of the UK market as subsidies are phased out; a comparison of participation in the wholesale market with co-located storage vs a long-term power purchase agreement with contracts for differences is made with help of a model using stochastic differential equations describing wind power production and prices.

Despite the abundance of literature in the matter, the application of high-scale, steady state  models, such as fluid queues, to energy storage remains relatively rare.
Limiting distributions for fluid queue models have been most commonly used in high-speed communication networks, but also for manufacturing \cite{mitra_stochastic_1988}, and energy systems, with early applications in hydro dam management models \cite{gaver_limiting_1962}. In \cite{oren_optimal_1986}, a two-state markov-modulated fluid queue (MMFQ) is used for optimizing the size and management policy of the national strategic petroleum reserve (SPR). 

More recently, there have been application of fluid queue models to renewable power plants with BESS. In \cite{chen_power_2020}, a model-predictive control (MPC) algorithm is proposed for solving a joint storage configuration problem, i.e. sizing and managing storage, and the MMFQ framework is used to analyze the performance of the algorithm in terms of reliability of power availability. 
In \cite{deulkar_sizing_2019}, the authors consider a case in which there is no control over the charge/discharge rate (akin to a balancing policy), and use MMFQ to find a relation, in particular an asymptotic relation, between the battery size and the loss of load probability for a given grid configuration, so that the problem of finding the optimal size of storage to hit a target loss-of-load probability can be solved. 
Unlike in these cases, here we use the MMFQ framework in the objective function of our optimization model directly.

While short term operation of co-located storage can be optimized using dynamic programming techniques and accounting for current information on state of charge, wind forecast and prices, the optimal sizing of the co-located storage is based on long run average behavior of the production-storage system under an optimized stationary policy.  For this purpose we employ the spectral method for characterizing limiting distributions of $n$-dimensional MMFQ described in \cite{mitra_stochastic_1988,dshalalow_fluid_1997}. Algorithms that are more numerically stable have been developed \cite{sericola_finite_2001,sericola_fluid_1999}, but with a limitation of a single state with negative drift. However, this limitation is too restrictive for our setting since it   would correspond to a single state where batteries can be discharged.
In \cite{akar_infinite-_2004}, an algorithm is proposed for calculating limiting distributions of markov-modulated fluid queues without solving an eigenvalue problem. The algorithm has promising results in terms of stability to obtain results of a desired precision and makes no assumptions regarding the structure of the fluid queue, which is a promising option for general application models like ours.

In this paper, we use limiting distributions directly to obtain simple expressions of long-run profits of the system, suitable for use in a sensitivity analysis. In particular, we are interested in investigating the effect of some key parameters such as cost of storage and available contract prices  on the financial outcome of the project, the optimal storage size and the optimal contracting strategy of a wind power plant trying to maximize its profit.
This simple model would be of interest to a wind power plant operator and project developer for early-stage project feasibility analysis. It could also be of interest for policy designers, to evaluate the financial viability of storage projects for wind power plants with and without special incentives, e.g. by considering tentative subsidies in the form of reduced feed-in-tariffs for surplus generation conditional on participation in long-term fixed quantity contracts.

\section{Problem setting}

We consider the case of a WPP that participates in the wholesale electricity market and has access to a long-term forward market.
The WPP is also evaluating the construction of a collocated Battery Energy Storage System (BESS), which will be used to mitigate imbalance costs.
In our setting, the wind farm operator needs to answer two main questions: how much storage should be installed and how much energy should be sold in a long-term contract. To answer these questions, it is also relevant to determine a stationary management (charge/discharge) policy for the BESS. 
Further aspects of our setting are described in more detail in the paragraphs below.

We consider that the WPP is a price-taker in the long-term forward market, where it has access to three contracts:
\begin{enumerate}
    \item A long-term fixed quantity contract, where it can choose how much to sell for a fixed price in every period of the wholesale electricity market.
\item A pay-as-demanded forward contract, which can be used to cover any shortfall in generation with respect to the pledge in contract 1.
\item A pay-as-produced forward contract, which can be used to sell any excess generation with respect to the pledge in contract 1.
\end{enumerate}
In order to avoid arbitrage opportunities, contract 2 must have a higher price than 1 and contract 3 must have a lower price than contract 1.

This setting can be seen as a special case of a WPP that participates in the day-ahead market as described next.
The WPP can commit to generate any quantity up to its capacity. The producer has no control over the wind power output and the commitment is made before the actual output is known. 
An imbalance penalty is paid for the difference between the energy commitment and the actual energy delivered.
The case studied here is the special case where the commitment is the same for all periods, the imbalance penalty for shortfall is non-negative and constant, and the imbalance penalty for surplus is non-positive (i.e. it is a reward, not a penalty) and constant.

To mitigate the imbalance cost, the wind power producer can charge/discharge the BESS, so that the net output of the plant, i.e. that perceived by the market is the combined production of the wind farm and the BESS.
Having a co-located BESS will have the effect of firming the WPP's energy output, and thus giving it access to other markets (e.g. a capacity market). However, we don't take any other sources of revenue into consideration in this setting.

The scope of our work is a high-level analysis that could be of interest for early-stage project developers or policy makers. We are interested in looking at infinite-horizon average profits by considering the steady-state behavior of the model proposed and performing sensitivity analyses on a number of key parameters.
To be consistent with this approach, we do not consider the possibility of using storage for arbitrage, since arbitrage opportunities cannot exist in steady state in an efficient market.
This approach suggests the assumption of existence of a long-run equilibrium in the forward market, which depends on many factors that are not considered in our model, such as long-run demand behavior and a stabilization of the cost of new capacity (which in the case of storage, for instance, is actually  expected to continue declining for some years \cite{bolinger_utility-scale_2021}). We do not make any such strong assumptions. While such an equilibrium may be far from being reached, this model is a high-level analysis for which this coarse approximation of reality is sufficient.

We do not specify any particular battery technology. We include in our analysis the issues of charge/discharge conversion efficiency and energy dissipation. However, we do not include the degradation of the battery because of usage and aging in our model. It has been found that this can be an important characteristic to take into account in sizing studies \cite{loukatou_optimal_2021}, so this could be an interesting feature to add to future versions of our model, but is considered out of the scope of this paper.

\section{Model}
\label{sec:model}

\subsection{General description}

As mentioned before, the case we study here is a special case of a day-ahead model. To emphasize the flexibility of our model and lay the groundwork for a future extension considering variable prices, we describe first the model of the day-ahead market allowing for the possibility of variable prices, and then focus on the special case addressed in this paper.

\subsubsection{The market}
We consider a day-ahead market in which the WPP is a price taker, so it commits to produce an amount $q_t$ (in MWh) during market period $t$ at a price $p_t$ (in \$/MWh), which is known at the time the commitment is made. During period $t$, the actual wind power output of the farm is $w_t$ (in MWh), while $r_t$ is the amount energy injected into the BESS ($r_t<0$ if the energy is extracted), so that the net output of the plant is $w_t-r_t$.
The imbalance is $y_t=q_t-(w_t-r_t)$ and the imbalance penalty charged to the WPP is $\Xi_t=\Xi(y_t,p_t)$. We assume $\Xi(\cdot)$ to be a known, deterministic, time-invariant function of the imbalance and the energy price.

For the case with constant prices, i.e. the case of forward contracts, $p_t$ is constant and the imbalance is defined by (\ref{eq:imbalance}). Note that the negative sign in front of $\kappa'$ implies that the WPP is not penalized for excess injection, but, on the contrary, sells it on the forward market.  To avoid arbitrage opportunities, we must have $\kappa'<1$ and $\kappa>1$.

\begin{equation}
    \Xi(y_t,p_t) = \begin{cases}
    \Xi^+(y_t,p_t)=-\kappa'p_t y_t & \text{ if }y_t\ge 0\\
    \Xi^-(y_t,p_t)=\kappa p_t y_t & \text{ if }y_t< 0
    \end{cases}\label{eq:imbalance}
\end{equation}
\noindent, with $\kappa',\kappa\ge 0$.

\subsubsection{The BESS}
Let $\rho_c$ and $\rho_d$ be the conversion efficiencies of charge and discharge respectively. Thus, the round-trip efficiency is $\rho=\rho_c\rho_d$. The quantity $r_t$ is measured from the exterior of the BESS, so the energy effectively stored in the battery is $\rho_c r_t$ (for $r_t>0$) and the energy effectively extracted from the battery is $-r_t/\rho_d$ (for $r_t<0$).

The capacity (size) of the battery is $b$ (in MWh). In order to determine an optimal size, we need to model the cost of installing and operating the battery. We express this cost as an amortized cost $c$ in \$/(MWh$\cdot$h). By doing this, we assume that the battery is replaced at the end of its lifetime with the same capacity and at the same cost. It is common to break down the cost of storage into a cost for the energy storage capacity and a cost for the inversion capacity. In this model, we do not consider inversion capacity restrictions; this could be a feature to include in extensions of our model.

An important factor in BESS is energy dissipation, i.e. the proportion of energy stored in the BESS that is spontaneously lost without any charging or discharging performed. This is usually expressed as a fraction $\eta$ of stored energy per unit time. This is inconvenient for our fluid queue model, as described below. So instead, we model dissipation as a constant leakage.

\subsubsection{Objective}

The profit during time period $t$ is thus 
$\Pi_t=p_tq_t - \Xi_t\big(q_t-(w_t-r_t)\big)$. In the previous paragraphs, we have referred to $t$ as a period for ease of exposition, given its similarity  with standard electricity markets. However, we propose here a continuous-time model, so $t$ actually, and in all instances in the remainder of this document, refers to an instant, and, consequently, the quantities $q_t,w_t,r_t$ are powers (in MW). Note that the price $p_t$ is indeed in \$/MWh, so that our profit $\Pi$ is an instantaneous profit rate, in \$/h.

We are interested in the long-run average profit, as defined in  (\ref{eq:avgProfit}). The expectation is taken with respect to the stochastic process of interest here, ($w_t,p_t$), as described next.

\begin{equation}
    \Pi = \lim_{T\to\infty}\mathbb{E}\left[\frac{1}{T}\int_0^T\Pi_t dt\right] - cb \label{eq:avgProfit}
\end{equation}

\subsubsection{Sources of uncertainty}
We consider two sources of uncertainty in our model: wind power output and energy prices. This is done by considering $(w_t,p_t)$ as a joint continuous-time stochastic process. 
In particular, we model it as a continuous-time Markov chain (CTMC) with a discrete state space $\mathcal S = \mathcal W \times \mathcal P$, with $\mathcal W$ and $\mathcal P$ being the discrete state spaces of $w_t$ and $p_t$ respectively. This stylized model allows for a convenient formulation of limiting distributions, as described in the next subsection. 

There are several examples of modeling wind speed and wind power output as a Markov chain in the literature, with applications in simulation of wind data series \cite{lopes_use_2012,tang_improvements_2015,xie_non-homogeneous_2017} but also in long-run analysis \cite{deulkar_sizing_2019,chen_power_2020}. It has been found that simply performing a max likelihood parameter estimation can result in a very good approximation of the limiting distribution, but  that a key metric to obtain a more accurate model is the autocorrelation \cite{brokish_pitfalls_2009}. In \cite{brokish_pitfalls_2009}, in the context of co-located storage sizing for robust operation of microgrids, it is found that Markovian models with an autocorrelation that poorly reflects that of the original data series can lead to underestimation of necessary storage by as much as 50\%. Autocorrelation performance can be improved by increasing the order of the Markov chain, but this increases the size of the state space exponentially, so that only chains of second or third order are of practical relevance. In \cite{papaefthymiou_mcmc_2008}, a rolling-average method is proposed to obtain higher autocorrelation performance in the lower range (0-6h) without increasing the size of the state space. More recently, authors have proposed non-homogeneous Markov chains \cite{scholz_cyclic_2014,xie_non-homogeneous_2017} to more accurately reproduce the autocorrelation of real data series, with much better results replicating the daily behavior of wind (i.e. the autocorrelation around integer multiples of 24h). These results can also be leveraged to capture seasonal changes of longer duration in the wind distribution.
The expected range of optimal storage sizes should also inform the decision of how much autocorrelation needs to be captured by the model. If these sizes lie in the 0-6 hour range, capturing daily autocorrelation is less crucial than if storage is expected to be in the 24h+ range.

The discussion above refers only to the Markovian modelling of wind speed or wind power output and provides enough justification for our modelling of wind power output as a CTMC in the constant prices case that is addressed in this paper. We could find no references in the literature of modelling wind power production and electricity prices as a joint CTMC as our general framework proposes. This might be regarded as an acceptable approximation for the high-level steady-state analysis considered here, but some additional justification, including some empirical support, should be provided when developing that stage of the model. We leave a more thorough discussion of this for a future work in which the variable-price version of this model is developed. For the case considered in this paper, we discuss some empirical performance of the model in section \ref{sec:empirical}.

\subsubsection{Optimization model}
Putting the previous pieces together, we are interested in solving the infinite-horizon average profit optimization problem in (\ref{eq:optModel}).

\begin{maxi}|s|
{q_t,r_t,b}{\Pi}{}{}
\addConstraint{b\ge 0}
\addConstraint{q_t\in[0,W]\;\forall t}
\addConstraint{q_t,r_t,b\in \mathbb{R}\;\forall t}
\label{eq:optModel}
\end{maxi}

\noindent$W$ is the plant capacity (in MW). For simplicity, we are neglecting any ramping charge/discharge limits in this model. Furthermore, we intend here to model an electricity market, so we will restrict our analysis to stationary policies where the bid is a function of the price, i.e. $q_t=q(p_t)$. We can also write this as $q_t=q^s$ when $p_t=p^s$, for $s\in \mathcal S$.

Finally, we will also be interested in stationary charge-discharge policies $r_t=r(w_t,q_t,p_t)$, which we write $r_t=r(w_t,q_t,p_t) = r^s$ when $(w_t,q_t,p_t)=(w^s,q^s,p^s)$. We must note, however, that this definition must be overriden if the storage is empty or full as summarized below.

\begin{gather*}
r^s>0 \text{ \& storage full }\Rightarrow r_t=0\\
    r^s<0 \text{ \& storage empty } \Rightarrow r_t=0
    \end{gather*}

In this paper, we focus on cases where strategic storage of wind power generation is not attractive because prices and imbalance penalties are constant in time. For this case, it is known that the optimal (cost-minimizing) policy is a balancing policy, i.e. $r_t = q_t-w_t$  \cite{harsha_optimal_2015,bitar_bringing_2012}.

\subsection{Steady-state approach}

By the ergodicity of CTMC, and since we are restricting our analysis to stationary policies, we can express the long-run average profit of (\ref{eq:avgProfit}) in terms of limiting distributions as in (\ref{eq:avgProfitStates}). In words, the long-run average profit is the sum over all states of the income minus the imbalance penalty, for which there are two cases: if storage is available, the policy can be followed; if storage is not available (empty or full), the policy must be overriden.

\begin{align}
    \Pi = &\sum_{s\in \mathcal S}\Big(
    p^sq^s\pi^s - \psi^s\Xi^s(q^s-w^s) \nonumber\\
    &- (\pi^s-\psi^s)\Xi^s(q^s+r^s-w^s)
    \Big) -cb \nonumber \\
    = &\sum_{s\in \mathcal S}\Big(
    p^sq^s\pi^s - \psi^s\Xi^s(q^s-w^s) \Big) -cb \nonumber\\
    = &\sum_{s\in \mathcal S}\Big(
    p^sq^s\pi^s - \psi^s\kappa p^s(q^s-w^s)^+ \nonumber\\
    &+\psi^s\kappa' p^s(w^s-q^s)^+
    \Big) -cb
    \label{eq:avgProfitStates}
\end{align}

\noindent, where $\boldsymbol \pi = (\pi^s)_{s\in\mathcal S}$ is the limiting distribution of the CTMC $(w_t,p_t)$ and  $\boldsymbol \psi = (\psi^s)_{s\in\mathcal S}$ is the long-run probability of storage unavailability (empty or full). The optimal charge/discharge policy is balancing the output, so $q^s+r^s-w^s=0$ whenever storage is available, and the imbalance penalty in that case is zero, which justifies the second equality. The third equality is obtained by plugging in the definition of the imbalance in (\ref{eq:imbalance}).
$\boldsymbol \pi$ can be easily determined from the generator of the CTMC $(w_t,p_t)$. 
The long-run probability of unavailable storage $\boldsymbol \psi$ can be determined from some results of fluid queue theory, as shown next.

\subsubsection{Limiting distribution}
Our model corresponds to the model of a Markov-modulated fluid queue with finite buffer.
A characterization of the long-run distribution of this process can be obtained through a spectral analysis, which we overview next. This is based on the presentation in  \cite{dshalalow_fluid_1997}, with more details of the proofs available in  \cite{mitra_stochastic_1988}. For ease of exposition, in the following overview we omit the efficiency factor $\rho$.

Define $\mathbf r=[r^s]_{s\in\mathcal S}$, a vector with the discrete values taken by $r_t$, $\mathbf D=\mathrm{diag} (\mathbf r)$, a diagonal matrix with $\mathbf r$ in its diagonal. In the field of fluid queues, $\mathbf{r}$ is called the drift vector. We assume for now that $r^s\neq 0\forall s\in\mathcal S$. The special case with $r^s=0$ is considered at the end.
Let $\mathbf{Q}$ be the infinitessimal generator matrix of the CTMC and $\mathbf{F}$ the limiting distribution of the level state of the battery, i.e.: $F(x,s) = \lim_{t\to\infty}\Pro\Big(X_t\le x,(w_t,p_t)=s\Big)$, $\mathbf{F}(x)=[F(x,s)]_{s\in \mathcal{S}}$.
Then, it can be shown \cite{dshalalow_fluid_1997} that $\mathbf{F}$ satisfies the differential equation 
\begin{equation}
\frac{d \mathbf{F}}{dx}\mathbf{D} = \mathbf{F}\mathbf{Q} \label{eq:ode}    
\end{equation}

with boundary conditions
\begin{align*}
    F(0,s) = 0 &\text{ if }r^s>0\\
    F(b,s) = \pi^s &\text{ if }r^s<0
\end{align*}

A spectral solution to these equations can be obtained introducing generalized eigenvalues $\lambda$ and eigenvectors $\mathbf{u}$, so that $\lambda \mathbf{u}\mathbf{D}=\mathbf{u}\mathbf{Q}$. The general solution to (\ref{eq:ode}) takes then the form
\[
\mathbf{F}(x) = \sum_{i=1}^{|\mathcal S|} a_i\exp(\lambda_i x)\mathbf{u}^i
\]
\noindent, where the values of coefficients $a_i$ can be found by solving a linear system from the boundary conditions. The long-run probability of unavailable storage can then be found as 
\begin{equation}
    \boldsymbol\psi = \mathbf F(0) + \boldsymbol\pi - \mathbf F(b)
    \label{eq:psi}
\end{equation}

In this manner, the long-run probability of unavailable storage can be found for given $\mathbf r$ and $b$. Note, however, that the method requires solving a generalized eigenvalue problem, which has three implications of importance for our work. First, we cannot obtain a closed-form expression of (\ref{eq:psi}) in terms of $b$ and $\mathbf r$, so that numerical calculating approaches are necessary. Secondly, this function is not convex in general, which makes our optimization problem non-convex as well. Finally, the linear system posed by the boundary conditions can be very ill-conditioned because of the presence of both very large and very small eigenvalues, which is a major challenge for the method. We address this in more detail in the description of the algorithm.

\subsubsection{Cases with zero drift}
If $r^s=0$ for some state $s\in\mathcal{S}$, then $F(x,s)$ can be expressed as a linear combination of $F(x,j)$ for states $j:r^j\neq 0$. Thus, for these cases, the method described above is performed on a reduced system that includes only states $\{j:r^j\neq 0\}$. Then those values are used to find the distribution for the null states. Details are omitted here, but can be found in the Appendix of \cite{sericola_fluid_1999}.

\subsubsection{Dissipation}

We next briefly address the issue of modelling dissipation under our model. As mentioned before, it is standard to consider that energy dissipates from the battery at a rate that is a fix multiple of the current storage level. This would mean that an additional term should be included in the drift vector: $r_t = r^s - \eta x_t$, where $x_t$ is the current storage level in the battery (in MWh). The addition of this term makes the drift dependent on the storage level, so that the spectral analysis performed earlier would be no longer valid. To avoid this, we model dissipation as a state-independent constant leakage. So that $r_t = r^s-\eta b$.

This affects the optimality of the balance policy. Indeed, for sufficiently large $\eta$, it could be more profitable to sell surplus energy immediately than to store it and have a large proportion of that lost to dissipation.
To simplify our analysis, we make the following restriction: we make a sensitivity analysis on $\eta$ with $\kappa^\prime=0$ and a sensitivity analysis on $\kappa^\prime$ with $\eta=0$, so that the balance policy is optimal for all these situations.

\subsection{Units}
In the previous paragraphs, we defined all quantities in their appropriate physical units. However, it is more convenient and illustrative for the purposes of this work to express all quantities in per unit of power plant capacity and storage capacity, so we introduce:

\begin{align*}
\mathbf{w} &= \Tilde{\mathbf{w}} W &\quad
q &= \Tilde q W &\quad
    b &= \Tilde b W \\
    c &= \Tilde c p^{max} &\quad
    p &= \Tilde p p^{max} &\quad
    \mathbf r &= \Tilde{\mathbf r} b
\end{align*}

\noindent, so that $\Tilde{\mathbf{w}}, \Tilde q, \Tilde p\in[0,1]$, $\Tilde b$ is in hours of storage of full plant capacity, and $\Tilde{\mathbf r}$ is in units of [p.u. of $b$]/h, which is interpreted as the number of times that the total storage would be charged starting from an empty state in one hour at full plant capacity.
Equation (\ref{eq:avgProfitStates}) becomes:

\begin{align}
\frac{\Pi}{p^{max}W} =
&\sum_{s\in \mathcal S}\Big(
    \Tilde{p}^s\Tilde{q}^s\pi^s - \psi^s\kappa \Tilde{p}^s(\Tilde{q}^s-\Tilde{w}^s)^+ \nonumber\\
    &+\psi^s\kappa' \Tilde{p}^s(\Tilde{w}^s-\Tilde{q}^s)^+
    \Big) - \Tilde c \Tilde b
\end{align}

It is not hard to check that this change of units does not affect the spectral decomposition and hence the values of $\boldsymbol\psi$.
In this paper we are interested in the case where the price is constant, i.e. $|\mathcal P|=1$, $p^s=p^{max}\;\forall s\in\mathcal S$ and hence $\Tilde{p}^s=1$ for all $s$.
Finally, for readability and ease of notation, the tildes will be omitted in the remainder of this paper, but we will always refer to values in per unit.

With these unit changes and simplifications, the optimization model (\ref{eq:optModel}) is written as

\begin{maxi}|s|
{q,b}{
\sum_{s\in \mathcal S}\Big(
    {q}\pi^s - \psi^s\kappa ({q}-{w}^s)^+ 
    +\psi^s\kappa' ({w}^s-{q})^+
    \Big) - cb
}{}{}
\addConstraint{b\ge 0}
\addConstraint{q\in[0,1]}
\label{eq:optModelPerUnit}
\end{maxi}




\subsection{Solution algorithm}
\label{sec:algorithm}

There are two challenges in finding the optimal solution to our problem. 
First, the spectral method used for computing the objective function is numerically unstable for certain values of $\mathbf r$, for which the solution involves a very ill-conditioned matrix.
Using Matlab's symbolic toolbox allows working with a degree of precision that reduces this problem. However, this significantly increases the computation time of the objective function. This difficulty can be circumvented with heuristic approaches as described next.
First, a computation in normal precision is tried. If boundary conditions fail to be respected within a predefined tolerance, the computation is repeated using the symbolic toolbox. We impose some precision limitations to reduce execution time.

It was observed that the drift vectors that elicit this behavior are those that include entries with very small absolute value. In the algorithm proposed in  \cite{sericola_finite_2001}, to guarantee obtaining results within a desired precision, the author imposes the restriction that entries in the drift vector should have a known infimum, which is in line with the behavior observed here. We use the following heuristic: if numeric instability persists after the previous procedure, a lower threshold is applied to the $\mathbf{r}$ vector. Specifically, all entries with an absolute value relative to the maximum absolute entry less than a certain threshold $r_{inf}$ are forced either to 0 or to the threshold, whichever is closer. If the problem persists, the threshold is increased until a stable solution is found.

This restriction does not make our solution too limited or unrealistic, since real charge/discharge equipment often does have a minimum threshold for operation, because of inverter and transformer's limitations, so that working with $r_{inf}=0.1$ is reasonable.
A more transparent way of introducing this heuristic rule would be to include it in the constraints of the optimization problem. However, this would make our problem a non-linear mixed-integer program, further complicating it without much perceived benefit. The more informal heuristic described above is thus preferred.
In a strict sense, enforcing this limitation implies not using the optimal balancing policy. 
However, in the cases where this happens, we disregard the effect that using the suboptimal policy  has on the optimality of the result.

An interesting alternative to address this numerical difficulty is using the algorithm proposed by Akar in \cite{akar_infinite-_2004}. This is discussed in more detail in the following subsection.

The second difficulty is that we have a non-linear, non-convex function, so that using a gradient descent algorithm does not provide a guarantee of global optimality. For the scenario with constant prices, our search space only has a dimension of 2; it is thus reasonable to start exploring the search space by means of a grid, to then use the best candidate as starting point for a finite-difference descent algorithm, which is the algorithm that was implemented in our tests.

\subsubsection{Eigenvalue-free algorithm}

In \cite{akar_infinite-_2004}, Akar and Sohraby propose an algorithm that allows obtaining the limiting distribution of a Markov-modulated fluid queue without solving an eigenvalue problem, to get around the problem of ill-conditioned matrices present in the spectral solution. They exploit a decomposition of the limiting distribution in what they call a constant, stable and antistable subsystem. This decomposition allows posing boundary conditions as a linear system without exponentially growing terms, which is thus  better suited for numerical solution.
Although they propose it in the context of communication networks, their algorithm does not assume any special structure for the driving Markov process or the drift vector, and can thus be readily used for our framework as well.

We implemented this algorithm but obtained mixed results, which are presented in Section \ref{sec:results}. In consequence, although this method seems promising for a continuation of this project, in the results shown in this paper we use the heuristic approach described earlier.

\section{Results and discussion}
\label{sec:results}

\subsection{Model data estimation}

\subsubsection{Wind model}
To obtain the values of the generator matrix that defines the CTMC, the methodology outlined in \cite{papaefthymiou_mcmc_2008} is followed.
The methodology consists in first passing the wind power output through an averaging window of one hour, then discretizing the output with $N=15$ levels,  and finally performing a max-likelihood estimation on the resulting data sequence, i.e. finding the transition probabilities from counting transitions in the sequence.
As in \cite{chen_power_2020}, the wind data available in \cite{nrel_wind_2016} was used. 

\subsubsection{Reference parameters}

We perform sensitivity analyses on the other model parameters. Reference values are given in Table \ref{tab:referenceValues}.
We comment on the choice of the reference cost of storage in the following paragraphs.

\begin{table}[!t]
\renewcommand{\arraystretch}{1.3}
\caption{Reference values for sensitivity analyses}
    \label{tab:referenceValues}
\centering
    \begin{tabular}{c|c||c|c}
         Parameter & Value & Parameter & Value  \\ \hline
         $\kappa$ & 1.35 & $\rho_0$ & 0.95\\
         $\kappa^\prime$ & 0 & $\rho_1$ & 0.95\\
         $c_s$ & 0.005  & $\eta$ & 0 \\ \hline
    \end{tabular}
\end{table}

\subsubsection{Empirical validation}
\label{sec:empirical}

To provide an empirical reference for comparison with our model, we use the same data series to solve the ex-post sizing and contracting  optimization problem \ref{eq:empiricalProblem}.

\begin{maxi}|s|
{q,b}{
g(q,b)
}{}{}
\addConstraint{b\ge 0}
\addConstraint{q\in[0,1]}
\label{eq:empiricalProblem}
\end{maxi}

\noindent, where $g(q,b)$ is a function that computes the ex-post average profit over the time available in the data series for contract quantity $q$ and storage size $b$. The value of function $g$ is computed by the algorithm below, where we use the fact that the the balancing policy is known to be  optimal.

\begin{algorithmic}
\REQUIRE $q,b,\mathbf{w},\kappa,\kappa',\rho,\Delta$
\STATE $\mathbf{w}$\COMMENT{Wind power output data series}
\STATE $\Delta$\COMMENT{Wind power output data series sampling period}
\STATE $T\gets \mathtt{length}(\mathbf{w})$
\STATE $x_0 \gets 0$ \COMMENT{Initial state of charge}
\STATE $x,x_{prev}$ \COMMENT{Current, previous state of charge}
\STATE $\Pi$ \COMMENT{Average profit}
\STATE $\Pi \gets q\times T\times \Delta$
\FOR {$t=1$ \TO $T$}
\IF{ $x_{prev} + \Delta t(w_t-q) <0$}
\STATE $\mathtt{empty} \gets$ \TRUE \COMMENT{empty storage}
\ELSE[Not empty]
\STATE $\mathtt{empty} \gets$ \FALSE
\ENDIF
\IF{ $x_{prev} + \rho\Delta(w_t-q) > b$}
\STATE $\mathtt{full}\gets$ \TRUE \COMMENT{surplus}
\ELSE[No surplus]
\STATE $\mathtt{full}\gets$ \FALSE
\ENDIF
\IF{$\mathtt{full}$}
\STATE $\Pi \gets \Pi + \kappa'\max\{\Delta(w_t-q)-(b-x_{prev})/\rho,0\}$
\ENDIF
\IF{$\mathtt{empty}$}
\STATE $\Pi \gets \Pi - \kappa \max\{0,\Delta(q-w_t)-(x_{prev})\}$
\ENDIF
\IF{$w_t-q <0$}
\STATE $x\gets \max\{\min\{ x_{prev} + \Delta(w_t-q), b\}, 0\}$
\ELSE
\STATE $x \gets \max\{\min\{ x_{prev} + \rho\Delta(w_t-q), b\}, 0\}$
\ENDIF
\STATE $x_{prev}\gets x$
\ENDFOR
\STATE $\Pi \gets \frac{\Pi}{T\Delta}$
\RETURN $g(q,b)=\Pi$
\end{algorithmic}

The results of the model (\ref{eq:optModelPerUnit}) and (\ref{eq:empiricalProblem}) are shown in Figure \ref{fig:storageValue_Pi} and \ref{fig:storageValue_q}. The results themselves are discussed in the next paragraphs; in terms of model validation, it is worth mentioning that there is a good level of agreement between the curves given by the model and the empirical ex-post best. An interesting continuation of this work would be to perform a similar analysis using a non-homogeneous Markov chain to model the wind power output as in \cite{scholz_cyclic_2014,xie_non-homogeneous_2017}.

\subsection{Value of storage}

As a first step before doing sensitivity analyses, we are interested in observing the value of storage for the WPP in the setting described. To do this, we fix $b$ at different values and solve (\ref{eq:optModelPerUnit}) and (\ref{eq:empiricalProblem}) for $q$ only to  find the optimal commitment and corresponding profit.
The profit is compared to the profit that would be obtained from a feed-in-tariff contract with the same price. In the setting that we consider here, there is no access to markets other than energy, so this profit is an upper-bound reference for comparison.

\begin{figure*}[!t]
\centering
\subfloat[ ]{\includegraphics[width=2in]{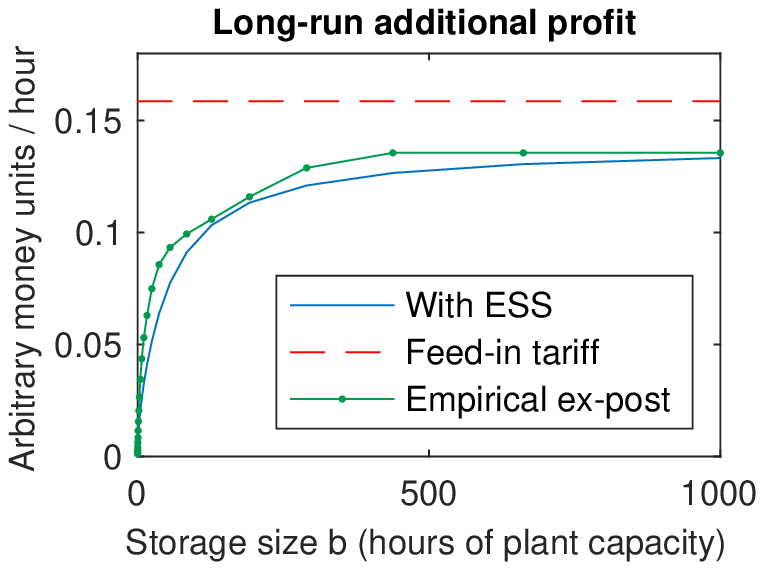}%
\label{fig:storageValue_Pi}}
\hfil
\subfloat[ ]{\includegraphics[width=2in]{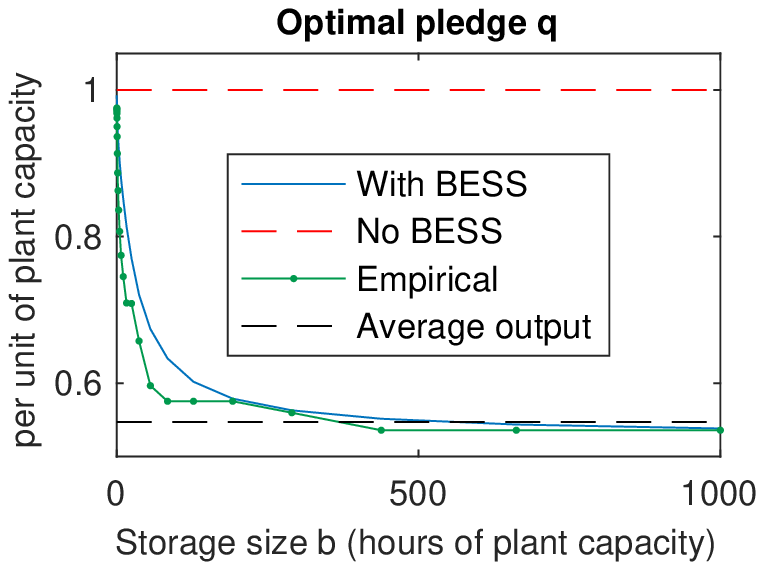}%
\label{fig:storageValue_q}}
\hfil
\subfloat[ ]{\includegraphics[width=2in]{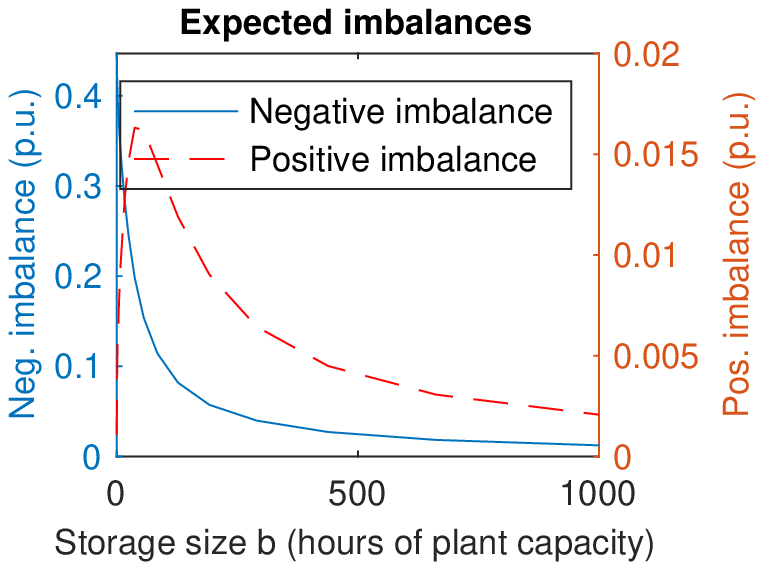}%
\label{fig:imbalances}}
\caption{(\ref{fig:storageValue_Pi}): Long-run average profit in excess of no-storage profit for different storage sizes. The feed-in-tariff benchmark is  equal to the average output of the power plant (price is normalized to 1). (\ref{fig:storageValue_q}): Optimal energy pledge for different storage sizes. (\ref{fig:storageValue_Pi}) and (\ref{fig:storageValue_q}) correspond to the reference values of all parameters.
(\ref{fig:imbalances}): Expected positive and negative imbalance for different storage sizes.} 
\label{fig:storageValue}
\end{figure*}

We can make four important observations from the results in Figure \ref{fig:storageValue}. First, as expected, even with unlimited storage, recovering feed-in-tariff profits is not possible, which is explained by efficiency losses in storage/discharge. Secondly, it is also worth mentioning that in our model, we do not find an optimal bidding function having the form of an optimal fractile of wind power production. Observe that since the support $\mathcal W$ is discrete, such a function would be step-shaped, while our optimal curve is not.

Third, in line with the results of other authors under different structural assumptions \cite{harsha_optimal_2015,bitar_bringing_2012}, the value of storage in the reference case studied has diminishing marginal profit. This allows identifying two quantities of interest. First, the marginal profit at $b=0$ gives the critical amortized cost of storage: if storage costs more than this quantity, it is not worth having co-located storage. The second quantity is the optimal size; for an amortized cost of storage $c$, the optimal size of storage is the quantity for which the marginal profit is equal to $c$. These results justify our choice of a descent algorithm to find the optimal storage size.

Finally, as expected, expected imbalances are reduced as storage sizes grows without limit. It is worth observing, however, that positive imbalances increase for the lowest range of storage sizes. This can be explained by the fact that in our model for the given reference values, the optimal contracting strategy in absence of storage is to commit the entire capacity of the plant, which eliminates positive imbalances. As storage drives the optimal commitment down, positive imbalances appear. With enough storage, these imbalances are then brought down too.

For the reference case shown here, the critical cost is found at $c_o=0.0193$, which for a reference price of energy of 60\$/MWh, corresponds to an amortized cost of 10.14\$/kWh-yr. Note that this is a whole order of magnitude below the reference cost of 100\$/kWh-yr for 2020 \cite{mongird_2020_2020}, meaning that under this setting and the reference values used, installing co-located storage is not profitable without access to additional revenue streams (such as a capacity market or arbitrage). Even taking into account that the cost of storage is declining rapidly, a tenfold reduction is beyond what we should expect to see in the near future. So additional revenue streams or a different setting need to be in place for co-located storage to be attractive.

\begin{figure}[!t]
\centering
\includegraphics[width=2in]{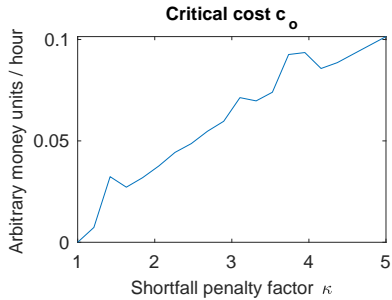}%
\caption{Critical storage cost for different values of shortfall penalty factor. Critical storage cost is the slope of the profit curve vs. storage size at $b=0$.}
\label{fig:criticalCost}
\end{figure}

In fact, in our setting, the WPP is not sufficiently exposed to uncertainty to justify paying for expensive storage. Indeed, the WPP is already covered against prices higher than $\kappa p$ by the pay-as-demanded forward contract, so that further risk hedging via storage is only modestly attractive. In Figure \ref{fig:criticalCost}, we see that as the coverage cap $\kappa$ is increased and the WPP is exposed to higher penalties, the critical cost $c_o$ at which it is willing to invest in storage also rises.
However, the pay-as-demanded forward contract would need to be available at a premium of several times the WPP's normal selling price to get anywhere near the critical cost.

A setting in which the WPP is exposed to much higher penalties is thus necessary for the co-located investment to be attractive, i.e. a setting with variable prices, which is the natural continuation of the work presented in this paper.

Finally, we obtain the identical curves in Figure \ref{fig:storageValue} using the spectral method described in Section \ref{sec:algorithm} and the alternative algorithm of \cite{akar_infinite-_2004}, with the latter having a much quicker execution. However, this agreement of results did not hold for some other values of the parameters, such that using that method for sensitivity analyses was not possible. Indeed, it seems that for some values of the parameters that were varied, the alternative algorithm can return invalid limiting distributions (outside $[0,1]$). This behavior has not been investigated any further but could be of interest for future extensions of the model.

\subsection{Sensitivity analyses}

As we mentioned, the main scope of the model proposed here is performing sensitivity analyses at a high level, such as during a feasibility study of a project, or for policy evaluation. In this section, we show the results of some of these analyses.
Despite the discussion of the previous section, sensitivity analyses  to changes in some key parameters are of interest, since they provide some understanding regarding the interplay between some key quantities which a project developer could be interested in evaluating in early stages of an investment. 

We are interested in observing the behavior of the model as some key parameters are modified. The current cost of storage, $c\approx 0.2$ is above the critical cost for the reference case, so the optimal size would be 0, and the behavior of the results to changes in other parameters would be hidden. To avoid this, we take a reference storage cost of $c=0.005$. Although unrealistically low, it allows observing the behavior of the optimal size as some parameters of interest change.
This could also be seen as the case of co-located storage with access to multiple revenue streams (e.g. capacity market), and only a portion of the full cost of the BESS needs to be recovered via risk mitigation in forward markets.

\subsubsection{Forward contract prices}
\begin{figure*}[!t]
\centering
\subfloat[ ]{\includegraphics[width=1.8in]{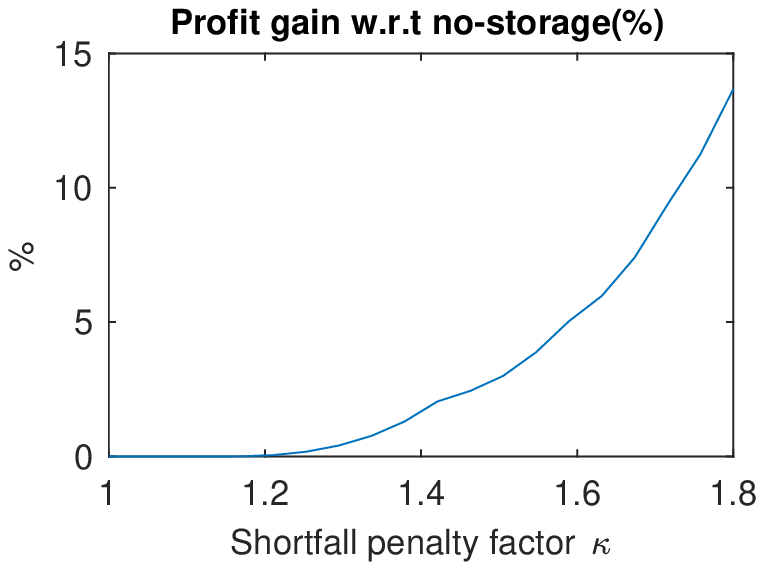}%
\label{fig:kappa_Pi}}
\hfil
\subfloat[ ]{\includegraphics[width=1.8in]{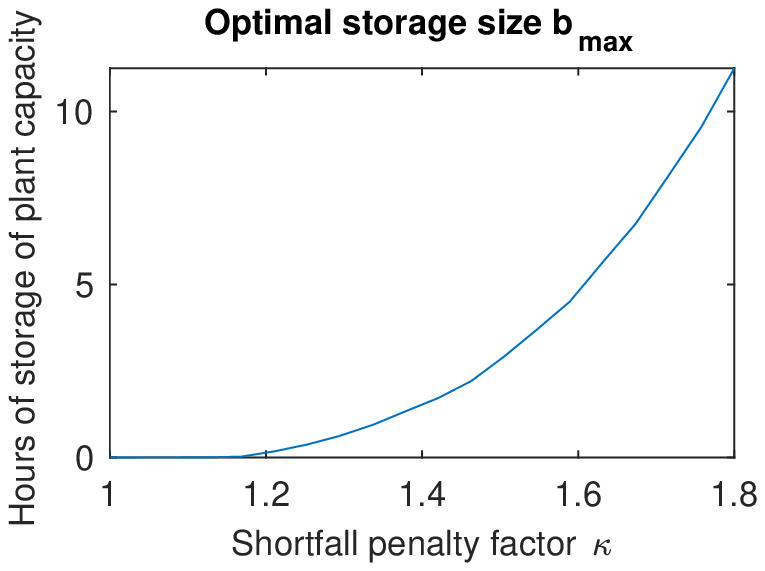}%
\label{fig:kappa_bmax}}
\hfil
\subfloat[ ]{\includegraphics[width=1.8in]{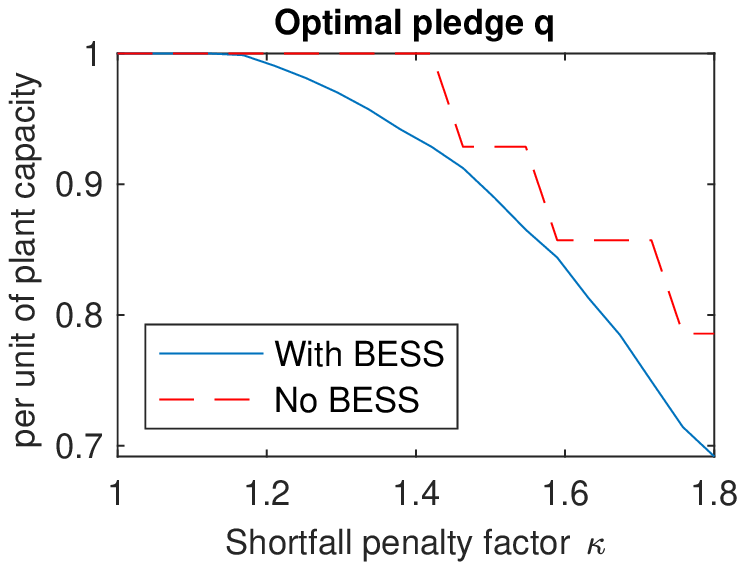}%
\label{fig:kappa_q}}
\caption{
Sensitivity to changes in shortfall penalty $\kappa$.
(\ref{fig:kappa_Pi}): Profit gain with respect to no-storage profit
(\ref{fig:kappa_bmax}): Optimal storage size
(\ref{fig:kappa_q}): Optimal quantity that should be sold in long-term fixed contract.
}
\label{fig:sensitivityKappa}
\end{figure*}

\begin{figure*}[!t]
\centering
\subfloat[ ]{\includegraphics[width=1.8in]{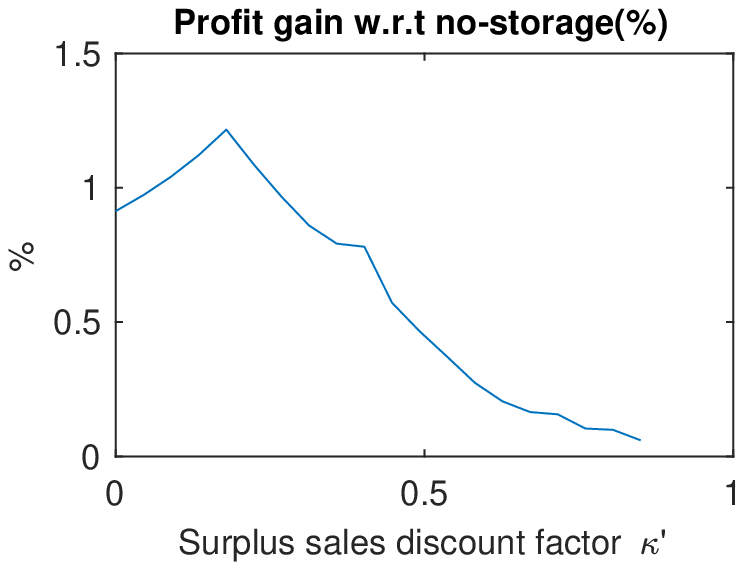}%
\label{fig:kappaPrime_Pi}}
\hfil
\subfloat[ ]{\includegraphics[width=1.8in]{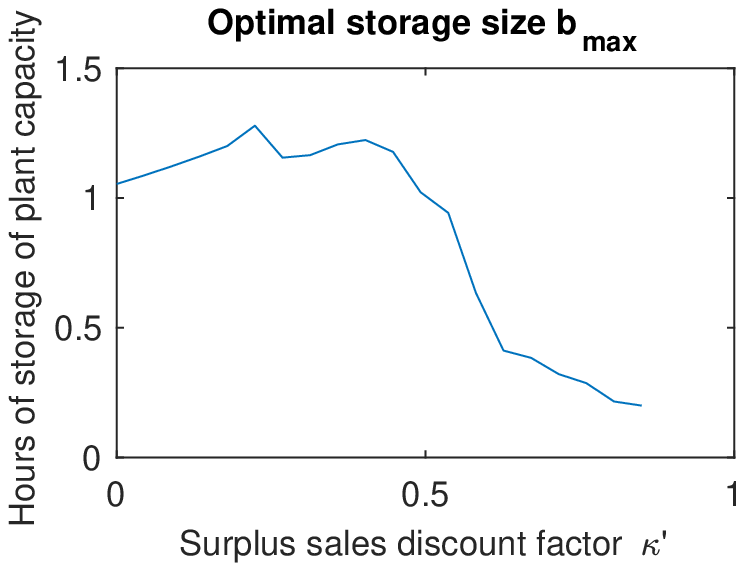}%
\label{fig:kappaPrime_bmax}}
\hfil
\subfloat[ ]{\includegraphics[width=1.8in]{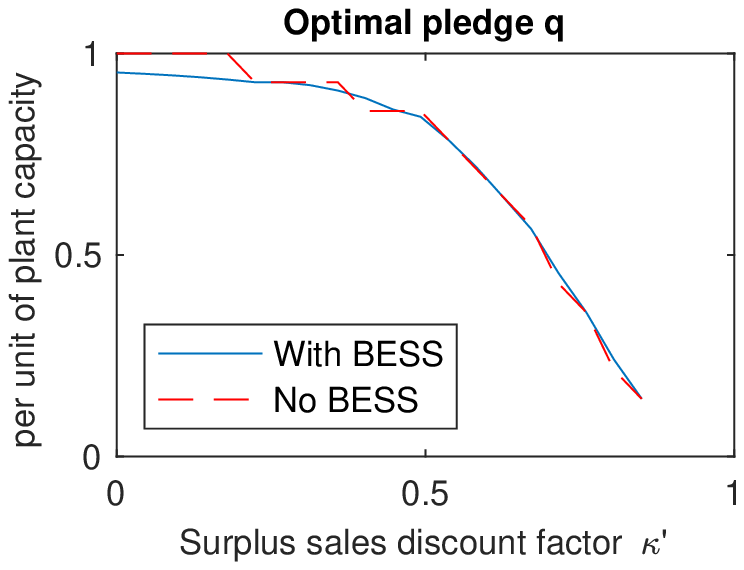}%
\label{fig:kappaPrime_q}}
\caption{
Sensitivity to changes in discount factor $\kappa'$ of surplus energy sales.
(\ref{fig:kappaPrime_Pi}): Profit gain with respect to no-storage profit
(\ref{fig:kappaPrime_bmax}): Optimal storage size
(\ref{fig:kappaPrime_q}): Optimal quantity that should be sold in long-term fixed contract.
}
\label{fig:sensitivityKappaPrime}
\end{figure*}

The behavior is exactly as intuition would dictate: larger storage becomes more attractive as the penalty for energy shortfall becomes higher, i.e. as the exposure to high penalties increases (Figure \ref{fig:sensitivityKappa}).
The change with respect to the price of the pay-as-generated forward contract (Figure \ref{fig:sensitivityKappaPrime}) shows a more interesting dynamic.
As $\kappa'$ becomes positive and the value of surplus energy increases, it becomes less undesirable to have excess energy, so that the optimal commitment $q$ decreases. This in turn leaves room for increasing the optimal storage size, as more energy is available for accumulation. After a certain point, however, the price of the pay-as-generated contract is so close to the price of the long-term fixed contract that the latter becomes less and less attractive, so that the commitment decreases significantly. With a very low commitment in the long-term contract, the motivation for investing in storage also disappears progressively, leading to a smaller optimal storage size.

\subsubsection{Efficiency and dissipation}
\begin{figure*}[!t]
\centering
\subfloat[ ]{\includegraphics[width=1.8in]{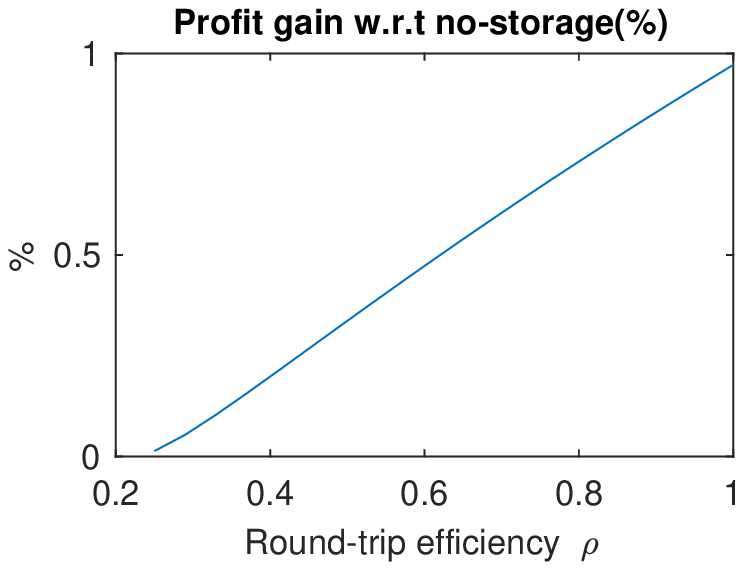}%
\label{fig:rho_Pi}}
\hfil
\subfloat[ ]{\includegraphics[width=1.8in]{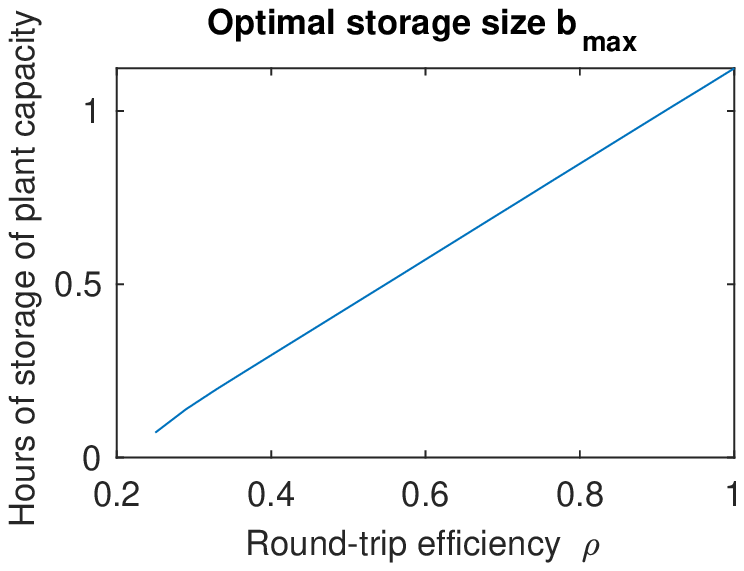}%
\label{fig:rho_bmax}}
\hfil
\subfloat[ ]{\includegraphics[width=1.8in]{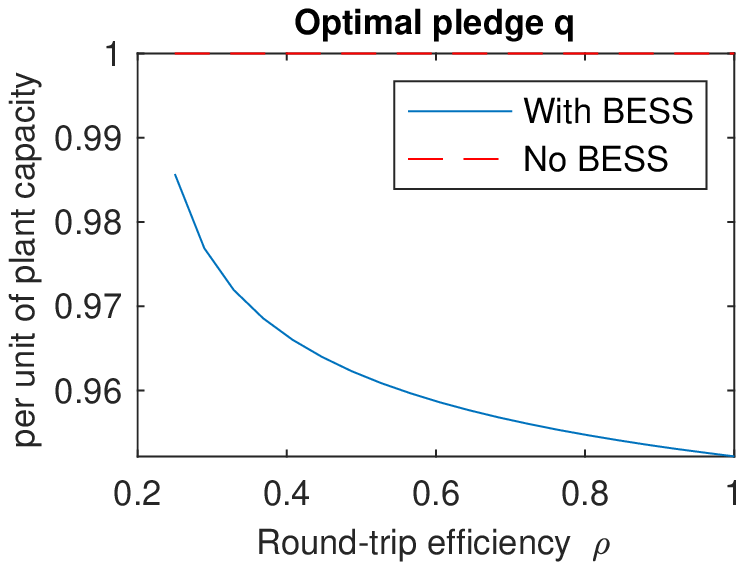}%
\label{fig:rho_q}}
\caption{
Sensitivity to changes in round-trip efficiency $\rho$.
(\ref{fig:rho_Pi}): Profit gain with respect to no-storage profit
(\ref{fig:rho_bmax}): Optimal storage size
(\ref{fig:rho_q}): Optimal quantity that should be sold in long-term fixed contract.
}
\label{fig:sensitivityrho}
\end{figure*}

\begin{figure*}[!t]
\centering
\subfloat[ ]{\includegraphics[width=1.8in]{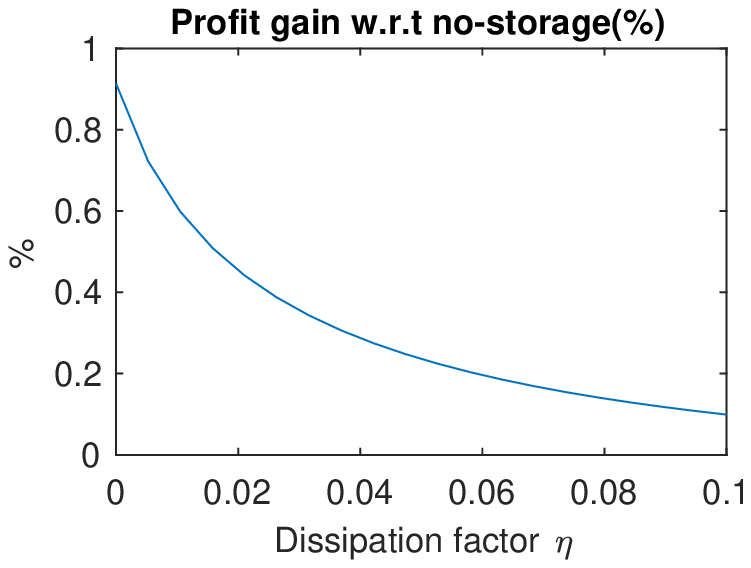}%
\label{fig:eta_Pi}}
\hfil
\subfloat[ ]{\includegraphics[width=1.8in]{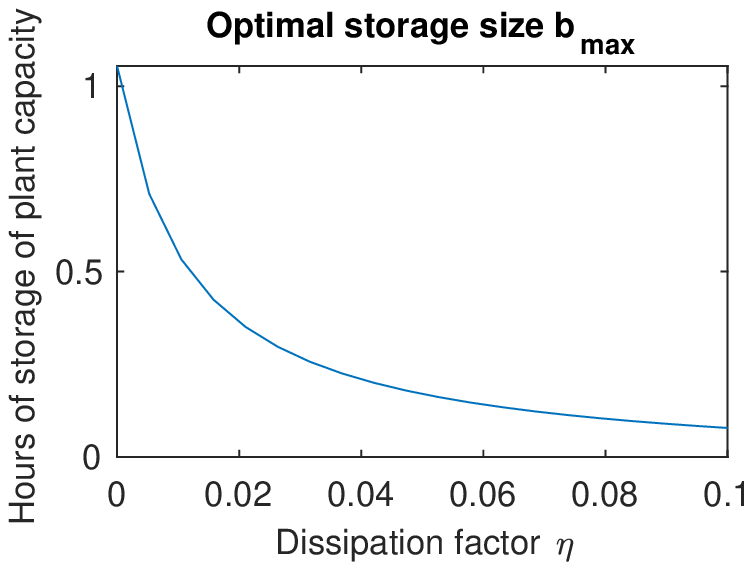}%
\label{fig:eta_bmax}}
\hfil
\subfloat[ ]{\includegraphics[width=1.8in]{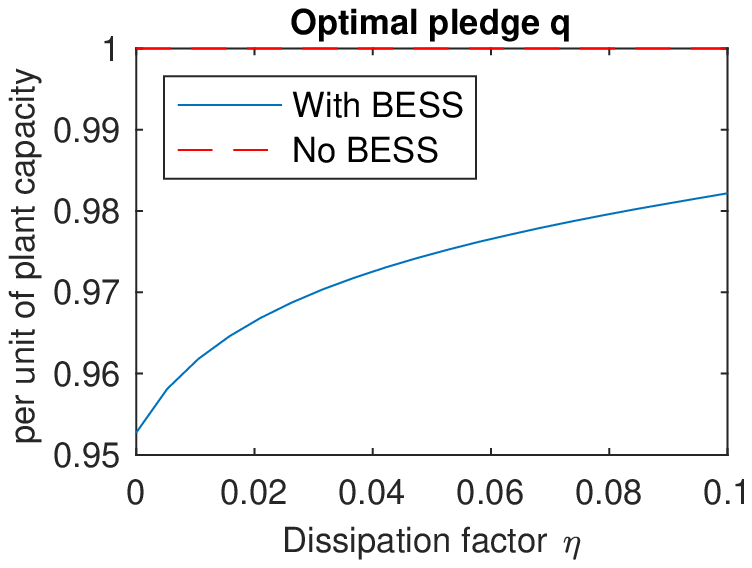}%
\label{fig:eta_q}}
\caption{
Sensitivity to changes in storage dissipation factor $\eta$.
(\ref{fig:eta_Pi}): Profit gain with respect to no-storage profit
(\ref{fig:eta_bmax}): Optimal storage size
(\ref{fig:eta_q}): Optimal quantity that should be sold in long-term fixed contract.
}
\label{fig:sensitivityEta}
\end{figure*}

The behavior when the round-trip efficiency and dissipation vary is in line with our expectation. As the system becomes more inefficient, storage becomes less attractive. Figure \ref{fig:sensitivityEta} shows this behavior as dissipation varies. For the round-trip efficiency parameter $\rho$, the relation between optimal size and efficiency appears to be close to  linear (Figure \ref{fig:sensitivityrho}).

\section{Conclusion and next steps}
In sum, we are proposing here a model to perform a high-level steady-state analysis of the value of co-located storage for a wind power producer that participates in the electricity market through long-term forward contracts.
In particular, we assess the optimal size of storage and optimal quantity to sell in forward contracts under different values of key parameters regarding contract prices and storage efficiency.
According to the model presented here, a setting such as the one considered in this paper does not expose the producer to enough uncertainty in income for storage to be attractive at current prices. This is likely to be different under a setting with variable prices, which is the natural extension of the model presented here, and what we intend to develop in future research.






\bibliographystyle{IEEEtran}
%

\bibliography{conference_references}

\end{document}